\documentstyle[amssymb,aps,prl,twocolumn,epsf,rotate]{revtex}

\begin{document}
\draft


\flushbottom
\twocolumn[\hsize\textwidth\columnwidth\hsize\csname
@twocolumnfalse\endcsname
\title{Mutagenesis and Metallic DNA}
\author{Chun-Min Chang$^1$, A.~H.~Castro Neto$^1$, and A.~R.~Bishop$^2$}

\address
{$^1$ Department of Physics,
University of California,
Riverside, CA 92521 \\
$^2$Theoretical Division and Center for Nonlinear Studies, Los
Alamos National Laboratory, Los Alamos, New Mexico 87545}
\date{\today}
\maketitle

\widetext\leftskip=1.9cm\rightskip=1.9cm\nointerlineskip\small
\begin{abstract}
\hspace*{2mm}
Positions of protons in DNA hydrogen bonds are fundamental for the fidelity
of the replication process. Because the hydrogen bond is partially
covalent the electronic structure of DNA plays an important role
in the replication process. Current
studies of electron transfer in DNA have shown that simple  pictures
based on independent electron physics are not sufficient to account for the
experimental data. In this paper we propose a model
for electron-proton interaction in DNA and its relationship to mutagenesis. 
We argue that
mutations will be more frequent in DNA sequences where electron 
delocalization occurs. To test 
our theory we calculate the heat capacity of DNA at low
temperatures and compare with the available experimental data in
hydrated DNA. We find that our model is in good agreement with the
data. We also propose new measurements of infrared absorption which
would further test our ideas.
\end{abstract}
] \narrowtext
\narrowtext

Watson and Crick \cite{watcri} pointed out the importance of proton transfer
as a possible cause of tautomeric base pairs in DNA and its deep
relationship to mutagenesis. Errors in the replication process of DNA can
lead to the occurrence of mutations related to genetic diseases and some
forms of cancer. L\"{o}wdin \cite{lowdin} argued that proton transfer is a
quantum mechanical effect which has purely electrostatic origin. Since then
many researchers have calculated the potential energy profiles for proton
transfer reactions in isolated base pairs \cite{clementi}. These studies
have shown the importance of the electronic configuration at each base pair
as a stabilizing factor of the hydrogen bond (H-bond). Indeed, recent
experiments \cite{zewail} show the occurrence of proton transfer in excited
base pairs. The problem of electron transfer in DNA has been a source of
intense debate \cite{debate}. Many different experimental studies of
electron transfer by chemical reaction have shown that long range electron
transfer in DNA may be rather different from the same process in proteins,
where single electron processes are dominant \cite{nunez}. In this paper we
propose a model which combines two essential ingredients in the physics of
DNA: the electronic degrees of freedom, and the protons in the H-bonds along
the double helix. Our model is different from other proposed models for
electron-proton interactions in DNA \cite{colson}, since we take into
account ({\it i}) the quantum mechanical character of electron motion, ({\it %
ii}) the charging energy due to electron transfer between different base
pairs, and ({\it iii}) the specificity of the electron-proton processes due
to differences in base pair sequences.

Let us consider a periodic DNA sequence. It is well-known that the electrons
in such DNA can move by the overlap of $\pi$ orbitals along the DNA helix 
\cite{piorb}. In this case DNA can be viewed as an one-dimensional (1D)
solid. Because a periodic DNA provides a periodic potential for electron
motion, basic quantum mechanics tells us that the quantum states of the
system are Bloch waves with wavenumber $k$ and band index $n$. To each value
of $k$ we associate a wavelength $\lambda$ of the electron ($k = 2
\pi/\lambda$), and for each band we associate a molecular or atomic state.
Band structure calculations \cite{band} indicate that such a DNA sequence
should be an insulator with a band gap of order of a few electron volts. In
this case the low energy excitations of the DNA are excitons. Recent
experiments in Poly(G)-Poly(C) DNA seem to confirm this picture \cite{porath}%
. However, electronic states in biologically relevant DNA (non-periodic)
depend considerably on the environmental conditions (such as pH, hydration,
etc). Although the study of periodic DNA sequences is very important, it
probably does not provide final information on the behavior of electrons in
DNA under conditions where mutagenesis is relevant.

Recent experiments on long range electron transfer in DNA indicate that the
introduction of Guanine-Citosine (G-C) base pairs enhance electron transfer 
\cite{gclong}. This situation is analogous to that in semiconductors where
the introduction of electron donors (acceptors) increases the charge
conduction by the creation of localized states close to the band edges. When
a sufficiently large number of donors or acceptors is introduced into the
DNA molecule an impurity band can be formed and conduction is possible. This
picture seems to be supported by recent measurements of DNA resistivity \cite
{resis}. Another possible source of 1D conduction is via non-linear
excitations such as solitons \cite{ahao}. As is well-known, disorder in 1D
systems leads to localization of electron wavefunctions \cite{anderson}.
However, when the localization length is large compared with typical
electronic scales, electron transfer occurs. Moreover, DNA is aperiodic
rather than random, and Anderson localization is a weaker effect\cite
{thierry}.

We assume from the outset that the charge carriers (electrons or solitons)
of DNA are delocalized over a region of size $L$ such that the energy level
spacing $\hbar v_{F,l}/L$ (where $v_{F,l}$ is the Fermi velocity in the
strand $l$) is much smaller than the thermal energy $k_B T$ (where $k_B$ is
the Boltzmann constant and $T$ is the temperature). The low energy
excitations in this case are particle-hole pairs close to the Fermi points
at $\pm k_{F,l}$. This simple scenario, however, is not enough to describe
charge propagation in DNA because the charging energies for electron
transfer between different base pairs is of order of electron volts as it is
common in organic systems. From this point of view DNA is very close to
being a Mott insulator instead of an ordinary band insulator \cite{anderson}%
. In 1D, such a conductor is called a Luttinger liquid \cite{luttinger}. In
a Luttinger liquid the low lying excitations are bosonic collective modes of
spin and charge which have solitonic character.

Long ago Pauling \cite{pauling} proposed that H-bonds have important
covalent character, which has subsequently been confirmed \cite{covalent}.
Proton states at the H-bonds are very important because they stabilize the
double helix. Protons are much heavier than the electrons and therefore much
slower. In A-T base pairs there are 2 H-bonds anchoring the two strands of
DNA, while in the G-C case there are 3 H-bonds. In principle each proton in
the H-bond can occupy one of its sides, and therefore there are 4 (8)
possible states for the protons in A-T (C-G) base pairs. These states do not
have the same energy because of the electrostatic interaction between the
atoms in the base pairs with the protons at the H-bonds. Moreover, the
difference in energy between the states is usually much larger than the
thermal energy and therefore we only need to consider the ground state and
the first excited state. We call the excitation energy between these two
states $\epsilon_{m}$, where $m$ labels the position of the base pair along
the DNA backbone. In this case the proton Hamiltonian can be simply written
as 
\begin{eqnarray}
H_p = - \frac{1}{2} \sum_m \epsilon_m \sigma^z_m  \label{hp}
\end{eqnarray}
where $\sigma^z$ is a Pauli matrix. In this representation the ground state
of the proton problem is the pseudo-spin state $|\Uparrow\rangle$ while the
excited state is $|\Downarrow\rangle$. Notice that $\epsilon_m$ can have
only two values corresponding to a A-T or G-C base pair. Thus, strictly
speaking $\epsilon_{m}$ is the energy required to produce a tautomeric state
on an isolated base pair. Because of the large proton mass we disregard
proton tunneling and treat the proton as a classical particle. Notice that
at temperature $T$ the probability of having a local proton configuration in
a mutated state is given by the Boltzmann distribution: 
\begin{eqnarray}
P_m(T) \propto \exp \left[-\epsilon_{m}/(k_B T)\right]  \label{pro}
\end{eqnarray}
corresponding to a thermal jump over the energy barrier. At room temperature
and with $\epsilon_m$ of the order of tenths of eV this probability is very
small and mutation is a rare event.

Let us now consider the coupling between protons and $\pi$ electrons. This
coupling depends strongly on the degrees of freedom of the H-bonds and the $%
\pi$ orbitals. On the one hand, the protons in the H-bond have their own
magnetic spin degrees of freedom and non-magnetic pseudo-spin variables
associated with their position in the H-bond. On the other hand, electron
states in DNA can be labeled by their momentum $\hbar k$, their true spin
magnetic quantum number, and by their two possible strand numbers: $l = \pm 1
$. While the strand and the spin quantum numbers of the electrons couple
directly to the pseudo-spin of the proton, the proton spin couples to the
electron spin via dipolar forces or contact interactions. If we view the
single H-bond problem as an impurity in a chain (or ladder) then the DNA
problem is of the so-called multi-channel Kondo type \cite{twochannel}.
Since in this paper we are only interested with the stability of the H-bond
due to the presence of electrons and not with the magnetic properties of
DNA, we neglect completely the dipolar interactions which can be studied
separately. Moreover, the symmetry of the coupling between pseudo-spins and
electrons is reduced due to the strong anisotropy of the proton positions.
Therefore, in the simplest model, the electron-proton problem reduces to a
local coupling of value $\lambda_{m,l}$ between the electronic charge
density and the H-bond at the base pair $m$ on the strand $l$: 
\begin{eqnarray}
H_{e-p} = \sum_{m,l} \lambda_{m,l} \left[\rho_{m,l}-\overline{\rho}_{m,l} %
\right] (\sigma^z_m-1)  \label{hep}
\end{eqnarray}
where $\overline{\rho}_{m,l}$ is the average electron density of the $m$th
base pair in the $l$th strand and $\rho_{m,l}$ the actual electronic
density. The physics of this problem is rather simple. In the absence of
electronic coupling the H-bond of a given base pair $m$ will find itself in
the lowest energy state which is stable at room temperature. Because the
proton is coupled to the electrons it can generate electronic displacements
along the helix with a gain of electrostatic energy. This gain of
electrostatic energy, in turn, can stabilize a tautomeric configuration of
the protons leading to mutation. The main question is whether the gain in
energy due to the electron-proton interaction is enough to compensate for
the loss of electrostatic energy between the proton and the base pair.
Moreover, the coupling between H-bonds and electrons also generates an
exchange interaction between H-bonds, which oscillates in space with a
wavelength of order of the distance $a$ between base pairs.

The low energy excitations of this Luttinger liquid can be studied by the
bosonization technique in which solitonic modes are bosons with momentum $%
\hbar k$, energy $\hbar v_{F,l}|k|$ and strand number $l=\pm 1$. In
Hamiltonian form it can be written as \cite{luttinger} 
\[
H_{L}=\sum_{k,l}\hbar \overline{v}_{F,l}|k|b_{k,l}^{\dag }b_{k,l}
\]
where $b_{k,l}$ ($b_{k,l}^{\dag }$) is the boson annihilation (creation)
operator and $\overline{v}_{F,l}=v_{F,l}\sqrt{1+2U_{l}/(\pi v_{F,l})}$ is
the renormalized Fermi velocity due to the electron-electron interaction
strength $U_{l}$. In our case the spin degrees of freedom are completely
decoupled (unless we introduce a direct coupling between the proton and the
electron spin) and do not appear explicitly into the problem. In particular,
the density operator which appears in (\ref{hep}) can be written as 
\[
\rho _{m,l}=\sum_{k}k\sqrt{\frac{L}{2\pi \left| k\right| }}e^{-\theta
_{l}}\left( b_{k,l}e^{ikx_{m}}+b_{-k,l}^{\dag }e^{-ikx_{m}}\right) 
\]
where $x_{m}$ is the position of the $m$th pair along the helix and $%
e^{\theta _{l}}=[1+U_{l}/(\pi v_{F,l})]^{1/4}$. The main gain in using the
bosonization procedure for this problem is that it maps the DNA problem
given by the Hamiltonian $H=H_{p}+H_{L}+H_{e-p}$ into an exactly solvable
problem in many-body physics called {\it the independent boson model} \cite
{mahan}. Our model describes the relaxation of the Luttinger liquid along
the helix due to the proton configuration.

The first result of the model is that the excitation energy $\epsilon_m$ of
the H-bond is reduced by an amount $\Delta_{m}$, where 
\begin{eqnarray}
\Delta_m = \frac{1}{2} \sum_{l=\pm 1} \frac{\lambda_{m,l}^2}{\pi v_{F,l} +
U_l} \, .  \label{only}
\end{eqnarray}
Thus, in the presence of a {\it local} metallic electron fluid one gets a 
{\it reduction} of the ``natural'' bias in the system and an increase in the
probability (\ref{pro}) of finding the system in a tautomeric state. It is
clear from (\ref{only}) that this can only happen when coupling between
H-bonds and electrons is large or when the Coulomb interaction between
electrons is weak. In metallic DNA the electron-electron interactions will
be screened by the electron motion and therefore $\Delta_m$ can be large
enough to induce mutation at room temperature. This mechanism for mutation
seems to be consistent with recent findings relating the existence of
repetitive sequences in DNA (which tend to delocalize the electrons) and
mutagenesis associated with evolution \cite{mutagen}.

In order to test our ideas we have calculated the heat capacity of DNA using
the above described model. There are three main contributors to the DNA
specific heat at low temperatures: the phonons which give a contribution
growing like $T^3$, the electrons which contribute with a term proportional
to $T$, and the protons which contribute with a Schottky term. The total
specific heat can be calculated once the DNA sequence is known. We assume
that the renormalized bias $\overline{\epsilon}_m = \epsilon_m - \Delta_m$
can be written as a series expansion of the temperature: $\overline{\epsilon}%
_m \approx A_m + B_m T$. The temperature dependence of the bias is due to
the exchange interaction between the H-bonds propagated by the electron
fluid. In a sample of DNA the values of $A_m$ and $B_m$ are distributed
around an average value which is given by the average number of G-C and A-T
pairs. For simplicity we assume here a gaussian distribution for both
quantities. In Fig. \ref{Cv} we compare the results of our model with the
experiments reported in Ref.\onlinecite{dnacold}. The agreement between our
model and the experiments is good. However, there are serious questions
about sample quality in experiments dealing with DNA and therefore we have
also calculated, as a further check, the infrared (IR) absorption for DNA
using many-body techniques. The correlation function of interest is the
proton spectral function which is defined as $S_m(\omega) = -2
Im[G_m(\omega)]$ where $G_m(\omega)$ is the Green's function of the proton
at base pair $m$ as a function of the frequency $\omega$. We found that IR
absorption at a certain site $m$ is controlled by a parameter $g_m$ defined
as 
\begin{eqnarray}
g_m = \frac{a \Delta_m}{\pi \hbar \overline{v}_F} \, .
\end{eqnarray}
At zero temperature and $g_m <1$, DNA shows an X-ray edge effect with a
sharp absorption edge. This effect is a result of the many-body character of
DNA electronic conduction. If $g_m>1$ or the temperature is finite the IR
absorption becomes smooth. For a specific DNA sequence one has to calculate
the absorption of the DNA as a whole. In Fig. \ref{xs} we show the
absorption spectrum for one of the DNA sequences of Fig. \ref{Cv} at finite
temperatures. Observe that no sharp X-ray edge is visible because of the
superposition of many X-ray edges coming from individual protons. There is,
however, a suppression of the absorption at low frequencies with the
formation of a ``pseudo-gap''.

In conclusion, we have proposed a model for electron-proton coupling and
mutagenesis in DNA based on the idea of charge delocalization. We have shown
that the position of the protons in DNA can change, leading to mutations
when the $\pi$ electrons delocalize. We have compared our results for the
specific heat at low temperatures with the available experimental data and
found near quantitative agreement. We have also proposed new IR absorption
experiments which can further test our model.

We acknowledge partial support from a Los Alamos CULAR grant under the 
auspices of the U.~S. Department of Energy.


\begin{figure}[th]
\caption{Specific heat of DNA : 1. DNA sample with water content $n_{0}=0-2$
M H$_{\text{2}}$O/MBP. The results are presented by the continuous curve
(theoretical) and $\circ $ (experimental). 2. DNA sample with water content $%
n_{s}=10-12$ M H$_{\text{2}}$O/MBP (mostly A-form DNA). The results are
presented by the dotted curve (theoretical) and $\Box $ (experimental). 3.
DNA sample with water content $n_{\Sigma }=22-23$ M H$_{\text{2}}$O/MBP
(mostly B-form DNA). The results are presented by the dashed curve
(theoretical) and $\triangle $ (experimental).}
\label{Cv}
\end{figure}


\begin{figure}[tbp]
\caption{Predicted infrared absorption at finite temperatures, $h\bar{v}%
_{F}/a=2%
\mathop{\rm eV}%
$ .}
\label{xs}
\end{figure}

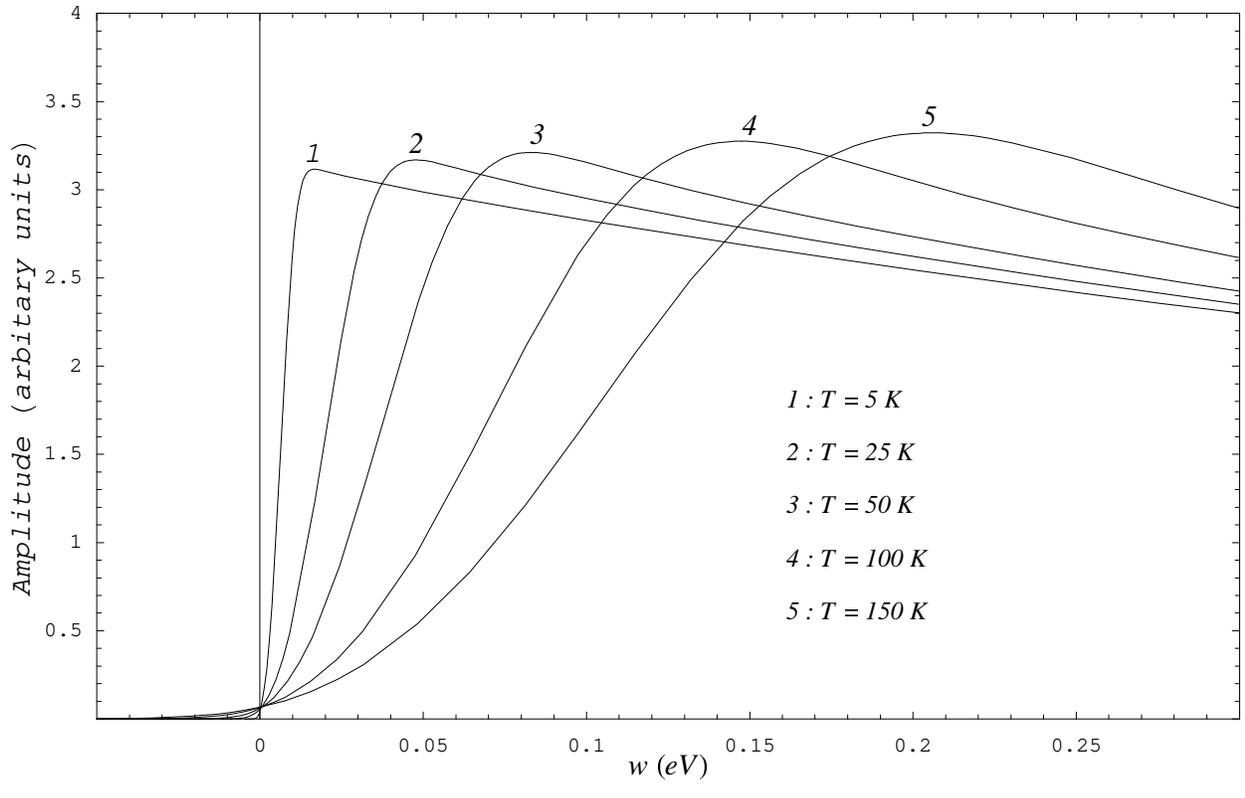
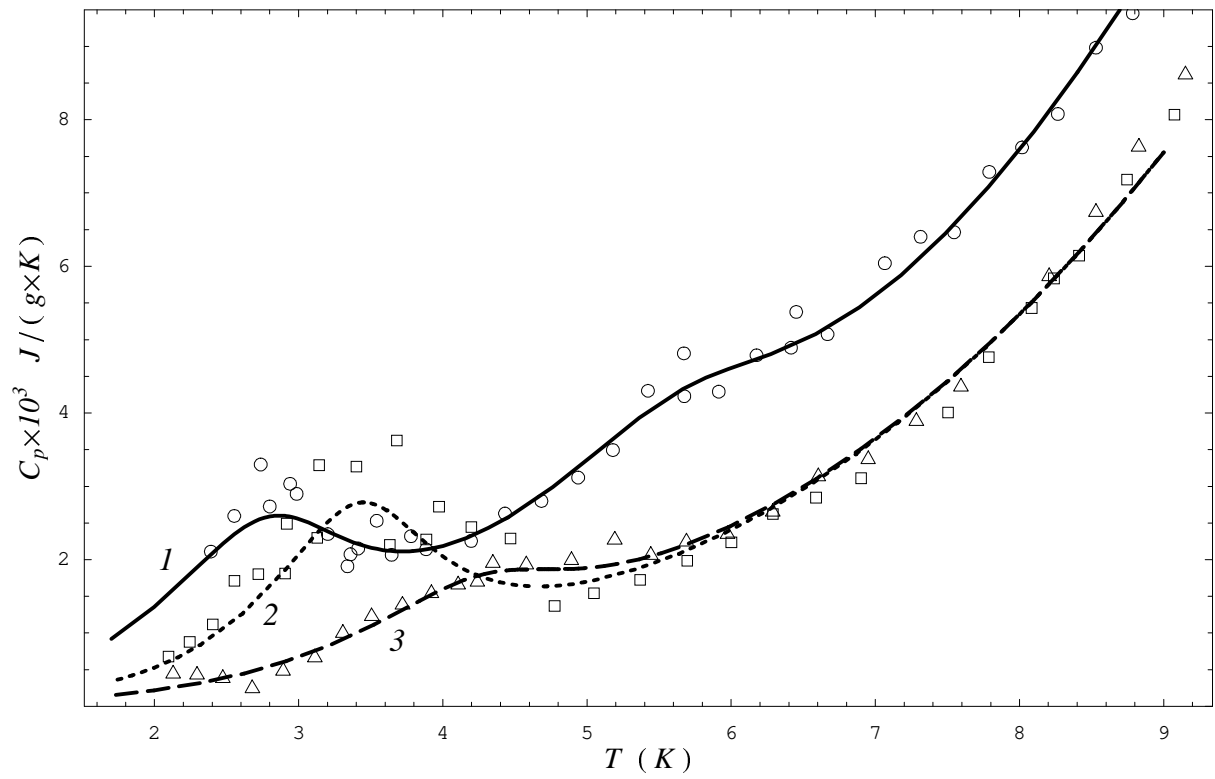
\end{document}